\documentclass[a4paper,oneside,11pt]{article}
\usepackage{cprform}
\usepackage{amsmath,amstext,amsfonts,amsbsy,amssymb,amscd,bbm,epsfig,lscape}

\usepackage{epsfig,macros,cite}
\usepackage{amssymb}
\usepackage{subfigure}

\newcommand{\parbreak}{\vskip 0.15cm}

\newcommand{\rep}[1]{\cite{#1}}

\newcommand{\ci}[1]{\boldsymbol{#1}}

\newcommand{\Dslash}{\relax{\kern+.25em / \kern-.70em D}}

\newcommand{\Real}{\relax{\mathsf{\Gamma\kern-.35em R}}}
\newcommand{\Int}{\relax{\mathsf{Z\kern-.40em Z}}}

\newcommand{\obar}[1]{\kern3pt\overline{\kern-2pt #1\kern-0pt}\kern1pt}

\newcommand{\corrbar}[1]{\kern3pt\overline{\kern-2pt #1\kern-0pt}\kern1pt}

\newcommand{\oVApAVren}[1]{\kern3pt\overline{\kern-2pt #1\kern-0pt}\kern1pt_{\rm\scriptscriptstyle VA+AV;s}}

\newcommand{\zbar}{\kern3pt\overline{\kern-2pt Z\kern-0pt}\kern1pt}

\newcommand{\zbarVApAV}[1]{\kern3pt\overline{\kern-2pt Z\kern-0pt}\kern1pt_{\rm\scriptscriptstyle VA+AV #1}}

\newcommand{\scrA}{{\rm\scriptscriptstyle A}}
\newcommand{\scrP}{{\rm\scriptscriptstyle P}}
\newcommand{\scrV}{{\rm\scriptscriptstyle V}}

\begin{document}
\bibliographystyle{mybibstyle}


\begin{titlepage}


\vspace*{-30truemm}
\begin{flushright}
ROM2F/2009/03\\
\vspace{5truemm}
\today
\end{flushright}
\vspace{5truemm}


\centerline{\Bigrm Quenched $B_{\rm K}$-parameter from Osterwalder-Seiler tmQCD quarks}
\centerline{\Bigrm and mass-splitting discretization effects}
\centerline{\Bigrm  }
\vskip 9 true mm
\begin{center}
\epsfig{figure=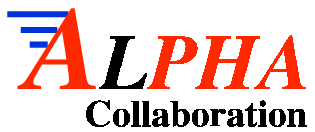, width=22 true mm}\\
\end{center}
\vskip -2 true mm
\centerline{\bigrm  P.~Dimopoulos$^a$, H.~Simma$^b$\footnote{On leave of absence from DESY, Platanenallee 6, D-15738 Zeuthen, Germany.} and A.~Vladikas$^c$}
\vskip 4 true mm
\centerline{\it $^a$ Dipartimento di Fisica, Universit\`a di Roma ``Tor
  Vergata''}
\centerline{\it Via della Ricerca Scientifica 1, I-00133 Rome, Italy}
\vskip 3 true mm
\centerline{\it $^b$ Dipartimento di Fisica, Universit\`a di Milano Bicocca,}
\centerline{\it Piazza della Scienza 3, I-20126 Milano, Italy}
\vskip 3 true mm
\centerline{\it $^c$ INFN, Sezione di ``Tor Vergata"}
\centerline{\it c/o Dipartimento di Fisica, Universit\`a di Roma ``Tor
  Vergata''}
\centerline{\it Via della Ricerca Scientifica 1, I-00133 Rome, Italy}

\vskip 10 true mm


\thicktablerule
\vskip 3 true mm
\noindent{\tenbf Abstract}
\vskip 1 true mm
\noindent
{\tenrm  We apply an Osterwalder-Seiler version of twisted mass QCD to a study of the $B_{\rm K}$ parameter, in which three of the four quark fields making up the relevant $\Delta S =2$ operator are maximally twisted with the same twist angle, while the fourth one has a twist angle of opposite sign. It is known that this setup ensures automatic improvement of the bare $K^0$-$\overline K^0$ operator matrix element and multiplicative renormalization of the  $\Delta S =2$ operator, at the price of breaking the $K^0$-$\overline K^0$ mass degeneracy by discretization effects. As a result, two dominant systematic errors of the $B_{\rm K}$ determination with Wilson fermions are kept under control. With the Clover term included in the fermion action, we perform a feasibility study and find, in the quenched approximation , a significant improvement of the scaling behaviour of $B_{\rm K}$, compared to earlier standard tmQCD determinations. Moreover, we study in detail the $K^0$-$\overline K^0$ mass splitting that characterizes this approach and confirm that, in the presence of the Clover term, it is greatly reduced in a maximally twisted theory.}
\vskip 3 true mm
\thicktablerule
\eject
\end{titlepage}

\section{Introduction}
\label{sec:intro}

The bag parameter of neutral $K$-meson oscillations $B_{\rm K}$, has been the object of many lattice QCD computations. Several discretizations of lattice fermions have been implemented for this purpose. For the most recent quenched result see ref.~\cite{Nakamura:2008xz}. Attention has now shifted to unquenched estimates; see ref.~\cite{Lellouch:latt08} for a recent review. 

The Wilson fermion results of $B_K$ are normally the least accurate, due to a limited control of two sources of systematic error:
\begin{enumerate}
\item
Loss of chiral symmetry causes the relevant $\Delta S = 2$ four-fermion operator to mix, under renormalization, with four other operators of the same dimension and different chiral representation \cite{
Bochicchio:1985xa}. In lattice discretizations which do not violate chiral symmetry this operator is multiplicatively renormalizable.
\item
Discretization effects of the $B_{\rm K}$ estimate at finite lattice spacing are $O(a)$; with staggered and Ginsparg-Wilson fermions they are $O(a^2)$. The traditional remedy of Symanzik improvement is not viable here, because of the large number of higher dimensional $O(a)$ counterterms required.
\end{enumerate}

The first of these problems has been dealt with in ref.~\cite{Becirevic:2000cy}, by using  Ward identities, and in refs.~\cite{Dimopoulos:2006dm,Dimopoulos:2007cn}, by implementing twisted Wilson fermions~\cite{Frezzotti:2000nk}. In the latter case, the lattice fermion action consists of a twisted up-down fermion doublet (tmQCD) and a standard  (untwisted) Wilson strange fermion. A second possibility, valid only in the quenched approximation, is that of a tmQCD  discretization of a degenerate down-strange doublet.
The lack of mixing with other operators in both tmQCD variants is achieved through the mapping of the usual, parity-even $\Delta S = 2$ operator to its parity-odd counterpart, which is known to be multiplicatively renormalizable, even with Wilson fermions~\cite{Bernard:1987pr,Donini:1999sf}. Both formalisms have been applied in refs.~\cite{Dimopoulos:2006dm,Dimopoulos:2007cn}, combined with the Schr\"odinger functional renormalization and RG-running of ref.~\cite{ssf:vapav}.

In order to implement $O(a)$ improvement, without the use of Symanzik counterterms, one must ensure that all flavours are regularized in a tmQCD framework at maximal twist. In other words, the twist angle for all quark flavours must be tuned to $\pm \pi/2$, while the mapping from parity-even to parity-odd $\Delta S = 2$ operator must be preserved. As discussed in ref.~\cite{tmqcd:DIrule} this is not possible in the standard tmQCD formalism. A way round this problem was provided in ref.~\cite{Frezzotti:2004wz}, by using somewhat different regularizations for sea and valence quarks. The former may be standard tmQCD, while the latter is the so-called Osterwalder-Seiler variant of tmQCD. Each valence quark is maximally twisted to $\pm \pi/2$. However, unlike in standard tmQCD, the valence flavours are not combined into isospin doublets.
The four-fermion operator consists of four distinct, maximally twisted valence flavours, three of which have twist angle with the same sign (say, $+\pi/2$) while the fourth twist angle has the opposite sign (say, $-\pi/2$). This combination of valence quarks ensures that the $\Delta S =2$ operator of interest is multiplicatively renormalizable and its $K^0 - \overline K^0$ matrix element is automatically improved; the same is true for $B_{\rm K}$.

It is not surprising that a price has to be paid for these advantages. In the setup described above, the $K^0$-meson consists of a strange/down valence quark-antiquark pair with, say, the same twist angle, while the $\overline K^0$-meson has a pair with opposite twist angles. This means that the two mesons have (pseudoscalar) masses which differ by $O(a^2)$ discretization effects. Although in principle this mass splitting vanishes in the continuum limit, it may be quite sizeable at finite lattice spacing. In the quenched approximation such effects have indeed been studied in the past (cf. refs~\cite{Jansen:2005cg,Abdel-Rehim:2006ve,Becirevic:2006ii}), being indirect manifestations of flavour symmetry breaking by tmQCD. 

In this work we have also performed a similar study of such flavour symmetry breaking effects for the $K$-meson mass $m_{\rm K}$ and decay constant $f_{\rm K}$. A comparison of our results with those of refs.~\cite{Jansen:2005cg,Abdel-Rehim:2006ve} indicates that the presence of the Clover term in the action greatly reduces such flavour breaking systematics, in accordance with the findings of an earlier study~\cite{Becirevic:2006ii}. Moreover, we find that the scaling of $B_{\rm K}$ with the lattice spacing is significantly improved, if compared to the previous standard tmQCD quenched studies of refs.~\cite{Dimopoulos:2006dm,Dimopoulos:2007cn}.
\section{Twisted and Osterwalder-Seiler valence quarks at maximal twist}
\label{sec:tmQCD-gen}

The setup of our formalism follows very closely that of  ref.~\cite{Frezzotti:2004wz}. Each valence
quark field $q_f$ (the subscript labels flavour, $q_f = u,d,s,\ldots$) is discretized by a fermionic action
of the form
\begin{align}
 \label{tmQCD_action2}
 S_f = a^4 \sum_x \,\,  \ {\overline q_f}(x) \big [ D_{\rm w} + m_f + i\gamma_5 \mu_f \big ]
 q_f(x) \,\,  ,
\end{align}
where $D_{\rm w}$ is the standard Wilson-Dirac fermion matrix {\it with a Clover term}. The Wilson plaquette action is the regularization of the pure gauge sector of the theory. In this work
all flavours are degenerate; this is only a matter of choice. As our simulations are performed in the quenched approximation, we will ignore issues concerning the lattice regularization of the sea quarks; see \cite{Frezzotti:2004wz} for a related discussion.

Let us consider quark bilinear operators with distinct flavours. We distinguish two cases, depending on the relative sign of the twisted mass terms (or equivalently, the twist angle) of these flavours. For example, the pseudoscalar density
\begin{align}
P_{ds}^{\rm tm}(x) \,\,\, = \,\,\, \bar d(x) \,\, \gamma_5 \,\, s(x) \qquad \qquad  ({\rm with \,\,  } \mu_d = - \mu_s)
 \end{align}
is said to be of ``twisted mass" (tm) type, if the corresponding twisted mass terms have opposite signs
(i.e. $\mu_d = - \mu_s$). In this case the two flavours may be grouped in a doublet, with a mass term of
the form $i\mu\gamma_5 \tau_3$, corresponding to the standard tmQCD formalism ($\tau_3$ is the Pauli isospin matrix). A second possibility is that  of $\mu_d = \mu_s$; this will be referred to as the
Osterwalder-Seiler (OS) case~\cite{Osterwalder:1977pc}. The pseudoscalar density is then denoted by
\begin{align}
P_{ds}^{\rm OS}(x) \,\,\, = \,\,\, \bar d(x) \,\, \gamma_5 \,\, s(x) \qquad  \qquad ({\rm with \,\, } \mu_d =  \mu_s) \,\, .
 \end{align}
Analogous definitions hold for other quark bilinear operators. The OS discretization is only applicable to valence quarks. Sea quarks, if regularized by the tmQCD lattice action, must be organized in doublets, with a flavour non-singlet twisted mass term $i\mu\gamma_5 \tau_3$, in order to avoid the generation of an unwanted $\theta$-term. The implementation of OS valence flavours is a mixed action formulation. Inevitably, in full (unquenched) QCD this introduces unitarity violation at finite lattice spacing.

The specific continuum operators of interest to us are $P_{ds}(x) = \bar d(x) \gamma_5 s(x)$ and
$A_{\mu,ds}(x) = \bar d(x) \gamma_\mu \gamma_5 s(x)$. As shown in ref.~\cite{Frezzotti:2000nk}, once the lattice theory is renormalized in a mass-independent renormalization scheme, such quantities are related, up to cutoff effects, to those of the lattice twisted theory through the quark field chiral rotations\footnote{Our notation is the following: the superscript ``cont" denotes continuum quantities, while $[ \cdots]_{\rm R}$
stands for lattice (re)normalized quantities, corresponding to either tm or OS discretization.}
\begin{equation}
\psi^{\rm cont}(x) =  \exp [ \frac{i}{2} \gamma_5 \tau_3 \alpha ] \,\, [\psi(x)]_{\rm R}
\qquad \qquad
\overline \psi^{\rm cont}(x)  =  [\overline \psi(x)]_{\rm R} \,\, \exp [ \frac{i}{2} \gamma_5 \tau_3 \alpha ] \,\, ;
\end{equation}
the phase $\alpha$ is the so-called twist-angle, defined through $\tan(\alpha) = \mu_{\rm R} / m_{\rm R}$. This of course refers to the standard tmQCD formulation, with an isospin quark doublet $\psi$ and
a twisted mass term of the form $i \mu \overline \psi \gamma_5 \tau_3 \psi$. For the case under consideration, with a distinct action~(\ref{tmQCD_action2}) for each flavour $q_f$, the corresponding rotations of the valence quark fields are
\begin{equation}
q_f(x)^{\rm cont}  =  \exp [ \frac{i}{2} \gamma_5 \alpha_f ] \,\, [q_f(x)]_{\rm R}
\qquad \qquad
\bar q_f(x)^{\rm cont}  =  [\bar q_f(x) ]_{\rm R} \,\, \exp [ \frac{i}{2} \gamma_5 \alpha_f ] \,\, ,
\label{eq:qf-rot}
\end{equation}
where the sign of the phase $\alpha_f$ is that of the corresponding mass term $\mu_f$. In order to ensure automatic improvement, we consider the case of maximal twist ($\alpha_f = \pm \pi/2$),
for which the above chiral rotations  induce the following relations for the two-fermion operators 
under consideration:
\begin{eqnarray}
P^{\rm cont}_{ds}  = & [P^{\rm tm}]_{\rm R}&  =  \hspace*{0.5cm} Z_P P^{\rm tm}_{ds}
\label{eq:Ptm}\\
A_{\nu,ds}^{\rm cont}  =  & \hspace{-0.25cm} - i [ V_{\nu,ds}^{\rm tm}]_{\rm R}&  =  -i Z_V  V_{\nu,ds}^{\rm tm}
\label{eq:Atm}\\
P^{\rm cont}_{ds}  =  & \hspace{-0.15cm}i [S^{\rm OS}]_{\rm R}&  =  \hspace*{0.35cm} i Z_S S^{\rm OS}_{ds}
\label{eq:POS}\\
A_{\nu,ds}^{\rm cont}  =  & [A_{\nu,ds}^{\rm OS}]_{\rm R}&  = \hspace*{0.5cm} Z_A  A_{\nu,ds}^{\rm OS} \,\, .
\label{eq:AOS}
\end{eqnarray}
All such relations, understood to be shorthand expressions for equations between correlation
functions (or matrix elements) involving these operators, are valid up to discretization effects.

Maximal twist is achieved by tuning the standard mass parameter $m_0$ to its critical value $m_{\rm cr}$ (with corresponding value $\kappa_{\rm cr}$  of the hopping parameter). There are several ways of doing this. A popular procedure consists in choosing a two-point correlation function which in the continuum violates parity and flavour symmetry and must therefore vanish. Thus, at fixed $\mu$, the point $m_0 = m_{\rm cr}(\mu)$ is determined, for which the bare correlation function vanishes. Finally, the critical mass  $m_{\rm cr}$ is found by extrapolating $m_{\rm cr}(\mu)$ to $\mu=0$ (unless the smallness of $\mu$ renders the extrapolation unnecessary). For details see ref.~\cite{Frezzotti:2005gi}. 
This procedure, based entirely on a tmQCD setup, ensures the absence  of $O(a^{2k})$ discretization effects (with $k$ integer) in the determination of $m_{\rm cr}$, while $O(a)$ effects are present. On the other hand, physical quantities like hadronic masses and matrix elements are automatically $O(a)$-improved; more precisely all $O(a^{2k+1})$ discretization effects are absent.

In the present work, we follow a very different approach, described in detail in refs.~\cite{Dimopoulos:2006dm, Dimopoulos:2007cn}. It consists in using the standard Schr\"odinger functional estimate of $m_{\rm cr}$, by requiring the vanishing of the PCAC quark mass, as in ref.~\cite{Capitani:1998mq}. This procedure uses standard (non-twisted) Wilson fermions,  thus obtaining an explicitly $\mu$-independent estimate of $m_{\rm cr}$. However, in order for this  $m_{\rm cr}$ to be $O(a)$-improved,
both the fermionic action and the axial current in the PCAC relation must have $O(a)$  Symanzik-counterterms. Therefore the non-perturbative estimates of $c_{\rm SW}$ and $c_A$, determined with standard Wilson fermions are essential in this procedure. With $m_{\rm cr}$ thus obtained in a non-twisted, improved Wilson fermion setup, we compute hadronic masses and matrix elements in a maximally twisted framework.
This approach has also been adopted in ref.~\cite{Becirevic:2006ii}, with lattices having periodic,
rather than Schr\"odinger functional, boundary conditions. In that work, a detailed comparison is made,
at $\beta = 6.0$, between pseudoscalar masses and decay constants (of the tm variety), obtained with the two estimates of $\kappa_{\rm cr}$ described above.

An important consequence, shared by these methods of tuning to maximal twist, is that the so-called ``chirally enhanced" discretization effects of the form $a^2/m_\pi^2$, present  in lattice correlation functions of fermion multi-local operators, are reduced to $a^2 \times a^2/m_\pi^2$; cf. ref.~ \cite{Frezzotti:2005gi}. This is however of little relevance to our results obtained close to the $K$-meson  mass region, i.e. far from the chiral limit.
\section{Results}
\label{sec:results}

We performed quenched simulations at three values of the gauge coupling, namely $\beta=6.0, ~6.1, ~6.2$. At each coupling we have tuned the degenerate quark masses $a\mu_d =  a\mu_s$ to a couple of values which are close to half the strange quark mass, so as to simulate charged K-mesons with degenerate flavours. In practice this entails tuning the twisted mass values so as to obtain pseudoscalar meson masses close to that of the physical Kaon. The values of $\kappa_{\rm cr}$, used in the simulations in order to achieve maximal twist, are collected in Table 3 of ref.~\cite{Dimopoulos:2007cn}. We use the renormalization constants, improvement coefficients and the ratio $r_0/a$ as collected in Appendix A of ref.~\cite{Dimopoulos:2006dm}. Physical quantities are usually expressed in units of $r_0$ in the present work~\cite{Sommer:1993ce}.

The details of our simulations are listed in Table~\ref{tab:runs}. The lattice size is $L^3 \times T$ with standard Schr\"odinger functional boundaries. The lattice calibration, expressed in terms of $a/2r_0$, is that of ref.~\cite{Dimopoulos:2006dm}. In the Table we also show the range (in units of $x_0/2r_0$) for which the ground state of our correlation functions (i.e. the K-meson) has been isolated. 
\begin{table}[!h]
\small
\begin{center}
\begin{tabular}{ccccccc}
\hline \hline
$\beta$   & $\frac{a}{2r_0}$  & (L,~ T)   &  $\big [\frac{x_0^{min}}{2r_0},~\frac{x_0^{max}}{2r_0}$ \big ] & $N_{\mbox {meas}}$   &  $a\mu_d = a\mu_s$ & dataset  \\ 
\hline
&&&&&&  \\
6.0       & 0.0931   & (24,~ 48) &   [1.30\,,~3.17]       &  100        &  0.0135  & I  \\ 
6.0       & 0.0931   & (24,~ 48) &   [1.30\,,~3.17]       &  100        &  0.0115  & II \\
&&&&&& \\ 
6.1       & 0.0789   & (24,~ 60) &   [1.26\,,~3.47]       &  100        &  0.0125  & I \\ 
6.1       & 0.0789   & (24,~ 60) &   [1.26\,,~3.47]       &  100        &  0.0110  & II\\
&&&&&&  \\
6.2       & 0.0677   & (32,~ 72) &   [1.35\,,~3.52]       &  50         &  0.0105  & I \\ 
6.2       & 0.0677   & (32,~ 72) &   [1.35\,,~3.52]       &  50         &  0.0090  & II \\
&&&&&&  \\
\hline \hline
\end{tabular} \vspace*{0.3cm}
\caption{Details of the runs for three values of the gauge coupling.}
\label{tab:runs}
\end{center}
\end{table}

\subsection{Comparison of discretization effects between tm and OS quantities}

All observables computed in this work are obtained from the large
time asymptotic limit of operator correlation functions with
Schr\"odinger functional boundary conditions.
The notation is standard, following closely that adopted e.g.
in  ref.~\rep{Dimopoulos:2006dm}. For instance, $f_\scrA$ ($f^\prime_\scrA$) denotes
the Schr\"odinger functional correlation function between a fermionic operator
$A_0$ in the bulk and a  pseudoscalar boundary
operator $\bar \zeta_s \gamma_5 \zeta_d$ at time-slice $x_0=0$ ($x_0=T$).
All such correlation functions are properly (anti)symmetrized in time,
when used to extract effective pseudoscalar masses
and decay constants. These quantities (as well as $B_{\rm K}$ in subsection~\ref{sub:BK})
are obtained at large time-separations from the boundaries, in order to avoid contamination
from higher excited states. Moreover, in this asymptotic limit, $O(a)$ discretization effects from
the boundaries may be considered negligible.
\begin{table}[!h]
\small
\begin{center}
\begin{tabular}{ccccc}
\hline \hline
$\beta$     &   dataset   &   $r_0M^{\rm tm}$   &  $r_0M^{\rm OS}$   &  $R_M$ in $\%$ \\
\hline
 6.0        &    I        &    1.255(5)      &   1.346(8)       &   15        \\
 6.0        &    II       &    1.169(5)      &   1.267(9)       &   17        \\  
 &&&& \\
 6.1        &    I        &    1.295(6)      &   1.349(8)       &   8       \\
 6.1        &    II       &    1.222(6)      &   1.279(8)       &   9       \\
 &&&& \\
 6.2        &    I        &    1.268(8)      &   1.301(10)      &   5       \\
 6.2        &    II       &    1.182(8)      &   1.217(10)      &   6       \\
&&&&  \\
\hline \hline
\end{tabular} \vspace*{0.3cm}
\caption{Pseudoscalar masses of tm and OS type and their mass splitting as defined in Eq~(\ref{Ratio}).  }
\label{tab:mK}
\end{center}
\end{table}

We have measured pseudoscalar effective masses $a M (x_0)$, both in the
tm and OS framework:
\begin{eqnarray}
a M^{\rm tm} (x_0) = \dfrac{1}{2} 
\ln \Bigg [ \dfrac{ f_{\scrV}^{\rm tm} (x_0-a)}{f_\scrV^{\rm tm} (x_0+a)} \Bigg ] \,\, ;
\qquad
a M^{\rm OS} (x_0) = \dfrac{1}{2} 
\ln \Bigg [ \dfrac{ f_{\scrA}^{\rm OS} (x_0-a)}{f_\scrA^{\rm OS} (x_0+a)} \Bigg ] 
\,\, .
\label{eq:effmass}
\end{eqnarray}
In Table~\ref{tab:mK} we show the results of the pseudoscalar meson masses from both regularisations.
Recall that the physical value of the Kaon mass in units of $r_0$ is $r_0 M_{\rm K} = 1.2544$. Therefore, 
for all three $\beta$-values, the K-meson decay constant $F_{\rm K}$ and the bag parameter $B_{\rm K}$  discussed below, are obtained through short ``interpolations" between two points.
We also show in Table~\ref{tab:mK} that the relative mass splitting, defined by
\begin{equation} \label{Ratio}
R_M =  \dfrac{(r_0 M^{\rm OS})^2 - (r_0 M^{\rm tm})^2}{(r_0 M^{\rm tm})^2} \,\,\, ,
\end{equation}
decreases with decreasing lattice spacing. This is a clear evidence that  the mass splitting is a discretisation effect. 

\vskip 0.0cm
\begin{figure}[!ht]
\subfigure[]{\includegraphics[scale=0.265,angle=-90]{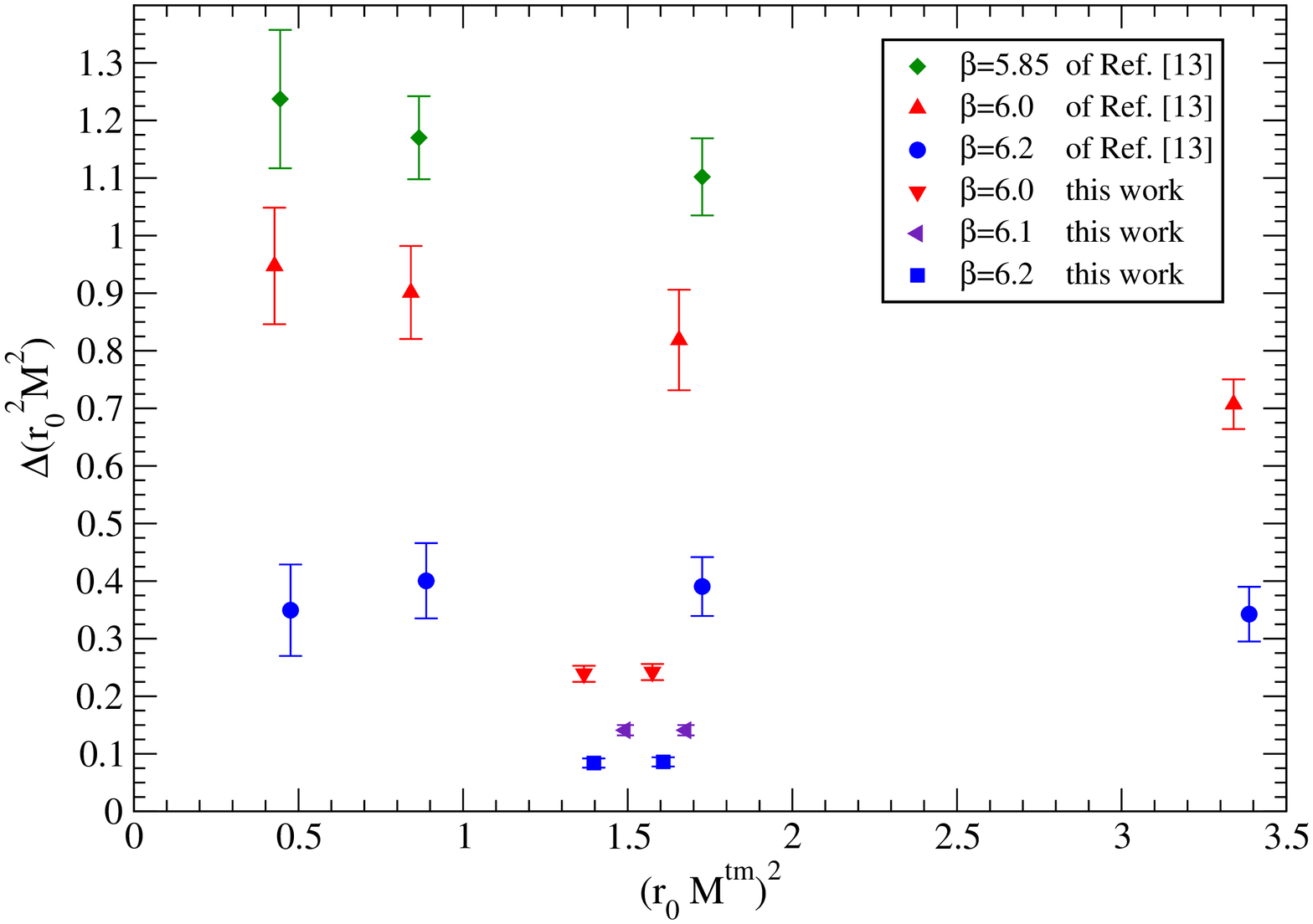}}
\subfigure[]{\includegraphics[scale=0.265,angle=-90]{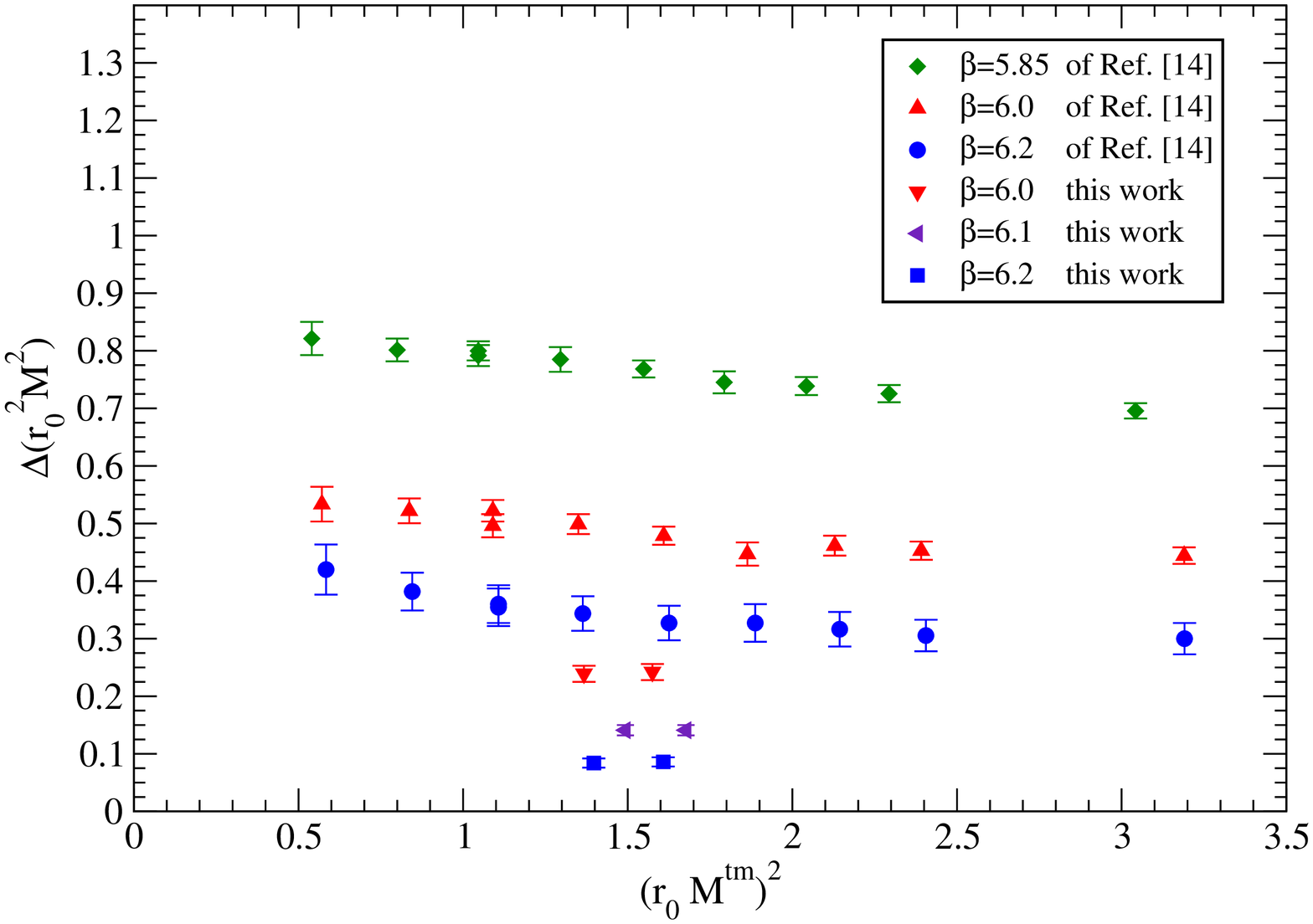}}
\vskip -0.5cm
\begin{center}
\caption{Square-mass splitting between tm and OS pseudoscalar mesons, as a function of the tm pseudoscalar meson mass: (a) Comparison between our results and those of ref.~\cite{Jansen:2005cg};
the latter are obtained using the PCAC determination of $\kappa_{\rm cr}$. (b) Comparison between our results and those of ref.~\cite{Abdel-Rehim:2006ve}.}
\label{fig:mass-split1}
\end{center}
\end{figure}
\vskip 0.0cm
\begin{figure}[!ht]
\begin{center}
\includegraphics[scale=0.45,angle=-90]{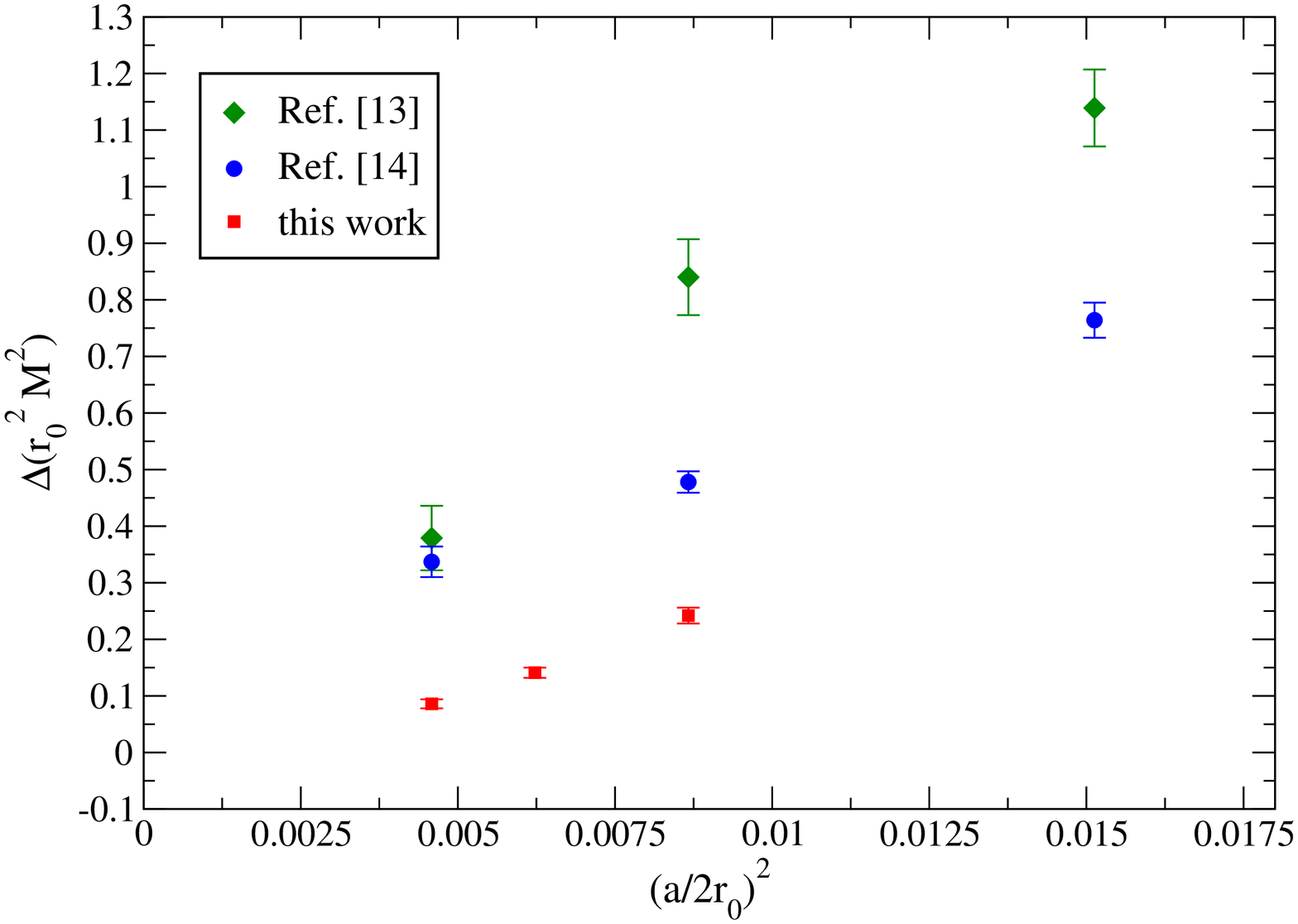}
\end{center}
\vskip -0.5cm
\begin{center}
\caption{Square-mass splitting between tm and OS $K$-mesons, as a function of the lattice spacing squared. 
The data of ref.~\cite{Jansen:2005cg}, displayed here,  are obtained using the PCAC determination of $\kappa_{\rm cr}$.}
\label{fig:mass-split2}
\end{center}
\end{figure}
In order to compare our results to those of other collaborations~\cite{Becirevic:2006ii,Jansen:2005cg,Abdel-Rehim:2006ve}, we consider the pseudoscalar meson mass splitting in units of $r_0$:
\begin{equation}
\Delta (r_0^2 M^2) \,\, \equiv \,\, [ r_0 M^{\rm OS} ]^2 - [ r_0 M^{\rm tm} ]^2 \,\,\, .
\end{equation}
The tm-OS pseudoscalar mass splitting may be affected by two factors: (i) the presence of the Clover term and (ii) the way  $\kappa_{\rm cr}$ is determined in order to ensure maximal twist. We first recall the different choices of fermion action and $\kappa_{\rm cr}$, made in refs.~\cite{Becirevic:2006ii,Jansen:2005cg,Abdel-Rehim:2006ve}:
\begin{itemize}
\item
Both refs.~\cite{Jansen:2005cg,Abdel-Rehim:2006ve} have data obtained with a maximally twisted Wilson action ({\it without a Clover term}) but implement different methods for computing $\kappa_{\rm cr}$. In particular, ref.~\cite{Jansen:2005cg} uses two methods, namely the PCAC relation for maximally twisted quarks (yielding an ``optimal" $\kappa_{\rm cr}$) and the vanishing of the pseudoscalar meson
mass in the standard (untwisted) theory. On the other hand the authors of ref.~\cite{Abdel-Rehim:2006ve} choose to tune  $\kappa_{\rm cr}$  through the vanishing of the parity-odd correlation function $\langle V^{\rm cont}(x) P^{\rm cont}(0) \rangle$, corresponding to another definition of an ``optimal" $\kappa_{\rm cr}$. A further difference is that in ref.~\cite{Abdel-Rehim:2006ve}, the values of $\kappa_{\rm cr}(\mu)$, computed at each twisted mass, are not extrapolated to $\mu \rightarrow 0$.
\item
In ref.~\cite{Becirevic:2006ii} the Clover term is included in the fermion action, just like in the present work. Their determination of $\kappa_{\rm cr}$ is based on the PCAC relation, evaluated in the standard Wilson theory (i.e. zero twist) with periodic boundary conditions, at large pseudoscalar masses, extrapolated to the chiral limit. On the other hand, our $\kappa_{\rm cr}$ determination  is based on the PCAC relation with Schr\"odinger Funtional boundary conditions, extrapolated  to the chiral limit from small masses\footnote{In our simulations, the twisted mass angle, measured from the ratio $\partial_0 f^{\rm tm}_V / \partial_0 f^{\rm tm}_A$, varies between $87^0 - 90^0$.}. Both estimates are $O(a)$-improved. 
\end{itemize}

Comparisons between results of these works are mostly limited to pseudoscalar mesons in the region of $M_K$. Pseudoscalar masses in the lighter mass range may be subject to finite volume effects (e.g. $L = 16^3$ at $\beta = 6.0$), while those related to heavier masses are subject to larger discretization errors. A close look at the details of these works suggests that the tm pseudoscalar masses in the $M_K$ range agree reasonably well, while the OS ones do not. Thus any discrepancies in $\Delta (r_0^2 M^2)$ are due to the latter mass. We may therefore conclude that these works indicate that the pseudoscalar mass $M^{\rm tm}$ is {\it weakly} dependent on the presence or absence of a Clover term and/or the way maximally twist is implemented.

Next, we examine the influence of $\kappa_{\rm cr}$ in the evaluation of $M^{\rm OS}$ and consequently of $\Delta (r_0^2 M^2)$. Our conclusions may be summarized as follows:
\begin{enumerate}
\item
The results of ref.~\cite{Jansen:2005cg}, obtained with two choices of $\kappa_{\rm cr}$, indicate that, for the coarser lattices, these quantities are affected by the way $\kappa_{\rm cr}$ is determined.
\item
The comparison made in ref.~\cite{Abdel-Rehim:2006ve}, between their $M^{\rm OS}$ and those of ref.~\cite{Jansen:2005cg}, obtained with the PCAC determination of $\kappa_{\rm cr}$, also shows discrepancies (cf. Fig.~1 of ref.~\cite{Abdel-Rehim:2006ve}; see also our
Figs.~\ref{fig:mass-split1},~\ref{fig:mass-split2}). Again the choice of $\kappa_{\rm cr}$ affects  $M^{\rm OS}$ and $\Delta (r_0^2 M^2)$.
\item
We observe a better agreement between $\Delta (r_0^2 M^2)$ of ref.~\cite{Abdel-Rehim:2006ve} and that of ref.~\cite{Jansen:2005cg}, obtained with the pseudoscalar determination of $\kappa_{\rm cr}$.
\item
From Fig.~5 of ref.~\cite{Becirevic:2006ii}, we estimate that at the $K$-meson region, $\Delta (r_0^2 M^2) \sim 0.27$ (for $\beta = 6.0$), while in our computations we find $\Delta (r_0^2 M^2) \sim 0.25$.
\end{enumerate}
The good agreement of the latter estimates suggests that once the Clover term is included in the action and an $O(a)$-improved $\kappa_{\rm cr}$ is used to tune to maximal twist, the tm-OS pseudoscalar mass splitting is only mildly affected by the details of the $\kappa_{\rm cr}$ determination. Conversely, it appears that the $O(a)$ effects in the ``optimal" determination of $\kappa_{\rm cr}$ (with a Wilson action without a Clover term) may somehow induce large $O(a^2)$ effects in $M^{\rm OS}$ and consequently in $\Delta (r_0^2 M^2)$.

We now investigate the direct influence of the Clover term in the mass splitting $\Delta (r_0^2 M^2)$. In Fig.~\ref{fig:mass-split1} we plot our data against those of refs.~\cite{Jansen:2005cg,Abdel-Rehim:2006ve}. The mass splitting of our results is significantly smaller\footnote{In refs.~\cite{Jansen:2005cg,Abdel-Rehim:2006ve} the
pseudoscalar masses $a M^{\rm tm}$ and  $a M^{\rm OS}$ are tabulated. From
these we can easily compute $\Delta (r_0^2 M^2)$.
The error on the latter quantity has been read off from Fig.~2 of
ref.~\cite{Jansen:2005cg} and Fig.~3 of ref.~\cite{Abdel-Rehim:2006ve}.
This rough error estimate is adequate for the present qualitative
comparison with our data.}. 
We attribute this to the inclusion of the Clover term in the action. This effect, already seen in ref.~\cite{Becirevic:2006ii} at $\beta = 6.0$, is confirmed here for more lattice spacings. As the data of refs.~\cite{Jansen:2005cg,Abdel-Rehim:2006ve} clearly indicate, this splitting depends weakly upon the quark mass values. We expect a similarly weak mass dependence in our case.

In Fig.~\ref{fig:mass-split2} we plot $\Delta (r_0^2 M^2)$ against the lattice spacing squared, for the present work and for refs.~\cite{Jansen:2005cg,Abdel-Rehim:2006ve}. In all cases this cutoff effect is decreasing with the lattice spacing. Contrary to expectations, there is no clear evidence for its vanishing in the continuum limit\footnote{This contradicts the claim of ref.~\cite{Jansen:2005cg}, where the mass splitting is shown to vanish in the continuum, for lighter pseudoscalar massess. In this respect we wish to point out that the tm and OS pseudoscalars were computed in that work with much bigger errors. Moreover, the $L=16^3, \beta = 6.0$ results at the lightest values of $a\mu$ may not be free of finite size effects. Thus we consider the conclusion of that work on the mass splitting under consideration as not definitive.}.
This is probably due to the fact that we only have results at three lattice spacings for each case. Moreover, our lattices are probably too coarse to reveal the vanishing of  $\Delta (r_0^2 M^2)$ in the continuum limit\footnote{Most peculiarly, the vanishing of $\Delta (r_0^2 M^2)$ in the continuum appears to be supported by our data if plotted against $a^4$. The data is well fit either by a function of the form $C_0 + C_4 a^4$ or by one like $C_2 a^2 + C_4 a^4$.}. 

Finally, we point out that this mass splitting has been recently studied in an unquenched $N_f=2$ framework, with a maximally tmQCD action {\it without a Clover term}; see ref.~\cite{Dimopoulos:2008hb}. In this preliminary work, pronounced $R_M$ values of up to about $50\%$ have been reported.

A similar analysis is performed for the pseudoscalar meson decay constants. In the Schr\"odinger functional framework they are obtained from the axial current correlation functions $f_{\scrA_{\rm R}}$, properly normalized by the boundary-to-boundary correlation function $f_1$; see~\rep{mbar:pap2} for details. For the maximally twisted tm and OS cases under investigation, the specific expressions
for the decay constants $F^{\rm tm}$ and $F^{\rm OS}$ are (in the large-time asymptotic regime):
\begin{eqnarray}
F^{\rm tm}  &\approx& 2 (M^{\rm tm} L^3)^{-1/2} 
\exp[(x_0 - T/2)M^{\rm tm}] \dfrac{- i Z_V f_{\scrV}^{\rm tm}(x_0)}{\sqrt{f_1^{\rm tm}}}\,\, ,
\label{eq:dc-ud}
\\
F^{\rm OS} &\approx& 2 (M^{\rm OS} L^3)^{-1/2} \exp[(x_0 - T/2)M^{\rm OS}]
\dfrac{Z_A f_\scrA^{\rm OS}(x_0)}{\sqrt{f_1^{\rm OS}}}\ .
\label{eq:dc-sc}
\end{eqnarray}
In practice the above quantities are obtained in the $x_0$-range in which the
pseudoscalar effective masses have been extracted.
\parbreak
A further method for computing $F$ is based on the PCVC relation, expressed in terms of Schr\"odinger
functional correlation functions:
\begin{equation}
-M^{\rm tm} Z_V f_\scrV^{\rm tm} (x_0) = 2 i \mu f_{\scrP}^{\rm tm} (x_0)\ .
\label{eq:exctPCVC2}
\end{equation}
The corresponding decay constant is computed as
\begin{eqnarray}
F^{\rm PCVC}  &\approx& - 4 \dfrac{\mu}{M^{\rm tm}}
(M^{\rm tm} L^3)^{-1/2} \exp[(x_0 - T/2) M^{\rm tm}] 
\dfrac{f_{\scrP}^{\rm tm}(x_0)}{\sqrt{f_1^{\rm tm}}}\ .
\label{eq:dc-ud2}
\end{eqnarray}

In Table~\ref{tab:fK} we show our results for the pseudoscalar decay constants. These 
values are in a general agreement with the corresponding estimates in Tables 6 and 7 of 
ref.~\cite{Dimopoulos:2007cn}, but the errors are bigger, due to reduced statistics\footnote{The quantities $r_0 F^{\rm PCVC}$ and $r_0 F^{\rm tm}$ have also been computed in
ref.~\cite{Dimopoulos:2007cn}. However these computations had been performed at significantly 
heavier quark masses and extrapolated to the Kaon mass value. This accounts for the $4\%$
difference between these results and ours at $\beta = 6.0$; this difference decreases with increasing $\beta$.}. 
The principal observation is that $F^{\rm tm}$ and  $F^{\rm OS}$ are compatible within errors.

\begin{table}[!h]
\small
\begin{center}
\begin{tabular}{ccccc}
\hline \hline
$\beta$     &   dataset   &   $r_0F^{\rm PCVC}$   &  $r_0F^{\rm tm}$   & 
$r_0F^{\rm OS}$ \\
\hline
 6.0        &    I        &   0.407(7)     &  0.411(6)      &  0.412(12)       \\
 6.0        &    II       &   0.397(6)     &  0.401(6)      &  0.408(13)       \\  
            &  $r_0M_{\rm K}$    &   0.407(6)    &   0.411(6)     &   0.412(12)       \\
 &&&& \\
 6.1        &    I        &   0.415(8)    &  0.419(8)       &  0.410(9)       \\
 6.1        &    II       &   0.407(8)    &  0.411(8)       &  0.404(10)       \\
            &  $r_0M_{\rm K}$   &   0.411(8)    &   0.415(8)      &  0.407(10)       \\
 &&&& \\ 
 6.2        &    I        &   0.417(12)    &  0.420(11)     &  0.419(14)       \\
 6.2        &    II       &   0.408(11)    &  0.411(11)     &  0.412(14)       \\
            &  $r_0M_{\rm K}$   &   0.416(11)    &   0.419(11)    &   0.418(14)       \\
 &&&& \\
\hline \hline
\end{tabular} \vspace*{0.3cm}
\caption{Pseudoscalar decay constants obtained from PCVC, and the axial current of tm
and OS type. We also show the interpolations at the physical K-meson point.}
\label{tab:fK}
\end{center}
\end{table}

\subsection{Improved $B_{\rm K}$ parameter}
\label{sub:BK}

We now pass to the computation of the $B_{\rm K}$ parameter, on the lines of ref.~\cite{Frezzotti:2004wz}, which ensures that the bare $B_{\rm K}$ is automatically improved.
The four-fermion operator of interest is
\begin{equation}
Q_{VV+AA}^{\rm cont} = [ \bar s \gamma_\mu d ] [ \bar s \gamma_\mu d ]
+[ \bar s \gamma_\mu \gamma_5 d ] [ \bar s \gamma_\mu \gamma_5 d ]  \,\, .
\end{equation}
On the lattice, we discretize this operator as proposed in ref.~\cite{Frezzotti:2004wz};
i.e. it is expressed in terms of current products made of both tm and OS type, at maximal twist:
\begin{equation}
Q_{VA+AV}^{\rm tm-OS} = 
[ \bar s \gamma_\mu d ]^{\rm tm} [ \bar s \gamma_\mu \gamma_5 d ]^{\rm OS}
+ [ \bar s \gamma_\mu \gamma_5 d ]^{\rm tm} [ \bar s \gamma_\mu d ]^{\rm OS}  \,\, .
\label{eq:VA+AV}
\end{equation}
Using the chiral rotations of eq.~(\ref{eq:qf-rot}), we obtain the relation between the lattice and continuum operators, up to discretization effects:
\begin{equation}
Q_{VV+AA}^{\rm cont} = [Q_{VA+AV}^{\rm tm-OS}]_{\rm R} = Z_{VA+AV} Q_{VA+AV}^{\rm tm-OS}
\end{equation}
The main reason behind using this mixed tm-OS formalism is the fact that in this way the $B_{\rm K}$ matrix element, computed with (twisted) Wilson fermions, is both automatically improved and multiplicatively renormalizable.
\begin{figure}[!h]
\begin{center}
{\includegraphics[scale=0.45, angle=-90]{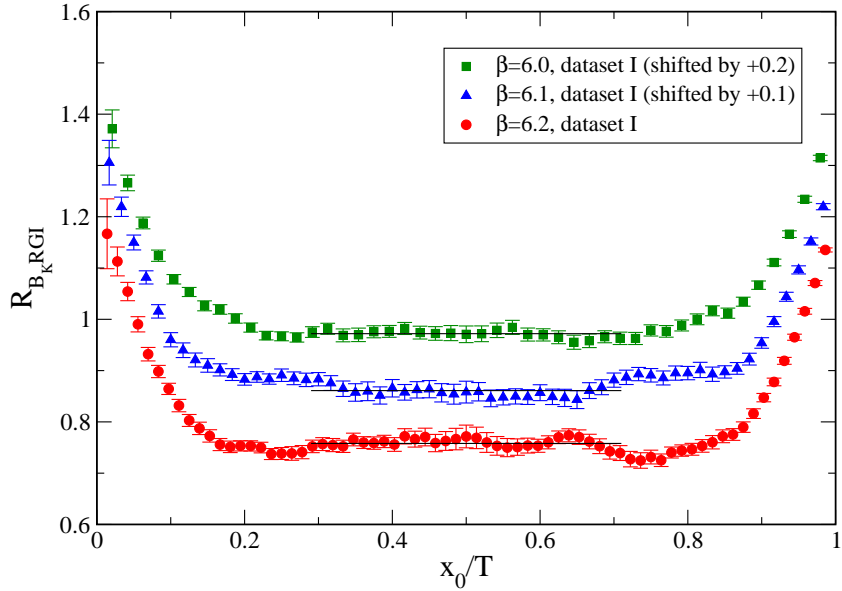}}
\caption[]{The ratio $R_{B_{K}^{\rm RGI}}$ as a function of time, for three values of the gauge coupling.
The data for $\beta=6.0$ and 6.1 have been shifted for clarity.  The horizontal lines indicate the $B_{\rm K}$ averages and the extension of the plateau.
}
\label{fig:BKplateau}
\end{center}
\end{figure}

$B_{\rm K}$ is obtained form the ratio
\begin{equation}
R_{B_{K}} = \dfrac{iZ_{VA+AV} F_{VA+AV}^{\rm tm-OS}}{(8/3) [iZ_V f_V^{\rm tm}] ~ [Z_A f_A^{'\rm OS}]} \,\, ,
\end{equation}
where $F_{VA+AV}^{\rm tm-OS}$ is the Schr\"odinger functional correlation function with 
the four-fermion operator of eq.~(\ref{eq:VA+AV}) in the bulk and the usual boundary sources
at the time edges. Note that these sources are of tm type at time-slice $x_0 = 0$ and of OS type at 
time-slice $x_0 = T$. Thus, away from the time boundaries, the correlation function $F_{VA+AV}^{\rm tm-OS}$ displays an asymptotic behaviour of the form $\exp[-M^{\rm tm} x_0] \exp[-M^{\rm OS} (T-x_0)]$. 
In order to match this choice and cancel the exponentials, the denominator consists of a correlation function $f_V^{\rm tm}$ of the tm type, involving the $x_0 = 0$ boundary, and a correlation function
$f_A^{'\rm OS}$ of the OS-type, involving the $x_0 = T$ boundary. The axial and vector currents in these correlation functions are the corresponding tm and OS twisted versions of the axial current; 
cf. eqs.~(\ref{eq:Atm}),~(\ref{eq:AOS}). In our analysis we have also considered the symmetric situation with tm and OS sources exchanged and the denominator of the above equation adjusted accordingly.
The $B_{\rm K}$ result quoted below is the average of the two estimates thus obtained. 
The quality of our raw results for $R_{B_{K}}$ is shown in  Fig.~\ref{fig:BKplateau}.

At large time separations from the boundary, the bare $B_{\rm K}$ estimate is automatically improved. 
The renormalization constant $Z_{VA+AV}$ is known non-perturbatively from ref.~\cite{ssf:vapav}. As it is computed in the chiral limit but is not improved, it suffers from $O(a \Lambda_{\rm QCD})$ discretization effects. These are not expected to dominate the scaling behaviour of $B_{\rm K}$; evidence for this will be provided below.

\begin{table}[!h]
\small
\begin{center}
\begin{tabular}{ccc}
\hline \hline
$\beta$     &   dataset   &   $B_{\rm K}^{\rm RGI}$  \\
\hline
 6.0        &    I        &   0.772(10)(9)  \\
 6.0        &    II       &   0.753(12)(9)   \\
            &  $r_0M_{\rm K}$   &   ~~~~~0.772(10)(9)(14) \\
 &&\\
 6.1        &    I        &   0.762(12)(9)  \\
 6.1        &    II       &   0.746(13)(9)  \\
             &  $r_0M_{\rm K}$    &   ~~~~~0.753(12)(9)(15)   \\
 &&\\
 6.2        &    I        &   0.758(12)(9)  \\
 6.2        &    II       &   0.742(13)(9)  \\
             &  $r_0M_{\rm K}$    &   ~~~~~0.756(12)(9)(15)  \\
 &&\\
\hline \hline
\end{tabular} \vspace*{0.3cm}
\caption{$B_{\rm K}^{\rm RGI}$ results of the present work. The errors are, in order
of appearance: (i) statistical, (ii) uncertainty due to the renormalization constants $Z_{VA+AV}$,
$Z_A$ and $Z_V$, (iii) total error, obtained by adding the first two in quadrature.
}
\label{tab:BK}
\end{center}
\end{table}

\begin{figure}[!h]
\begin{center}
\subfigure[]{\includegraphics[scale=0.248,angle=-90]{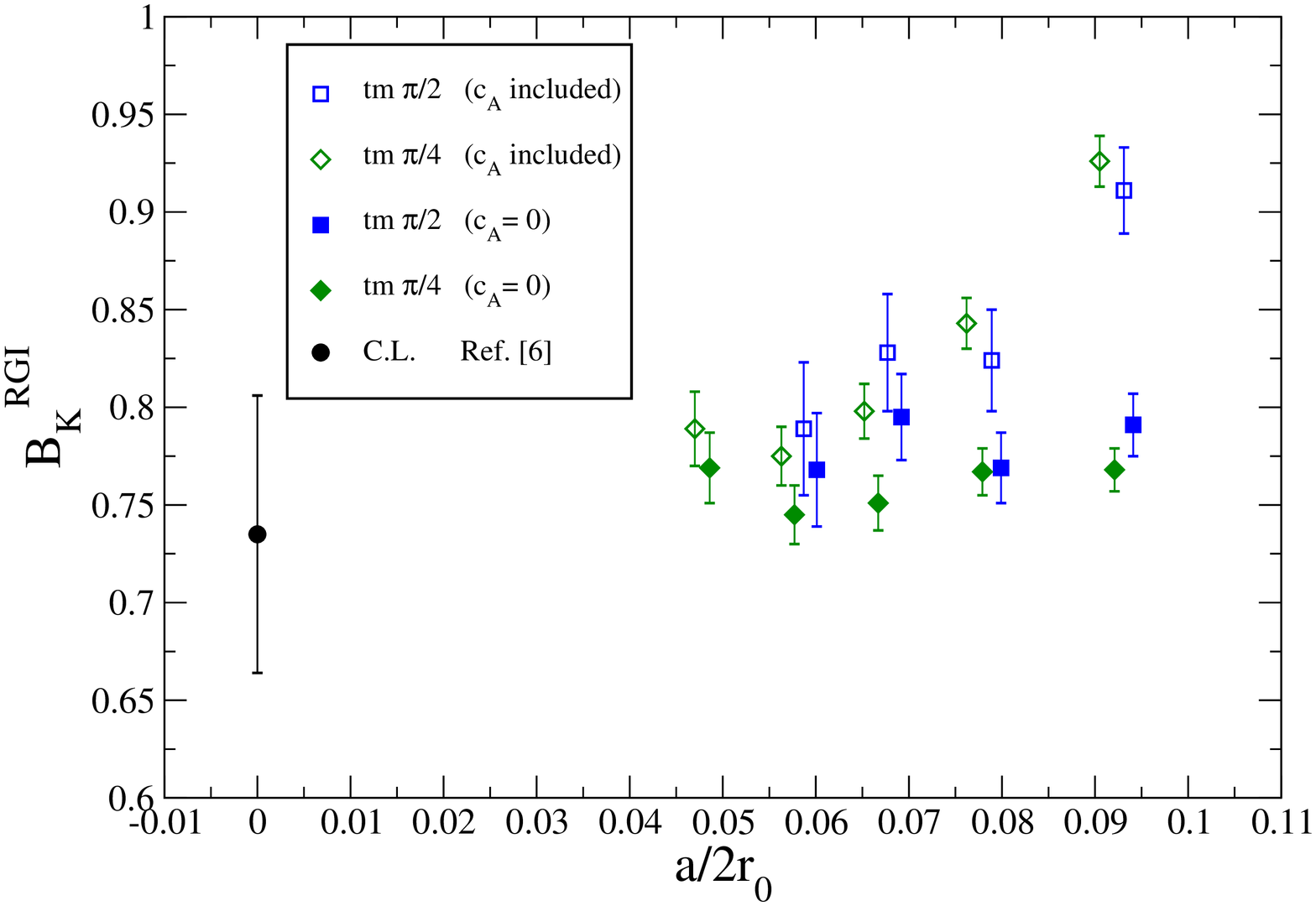}}
\subfigure[]{\includegraphics[scale=0.248,angle=-90]{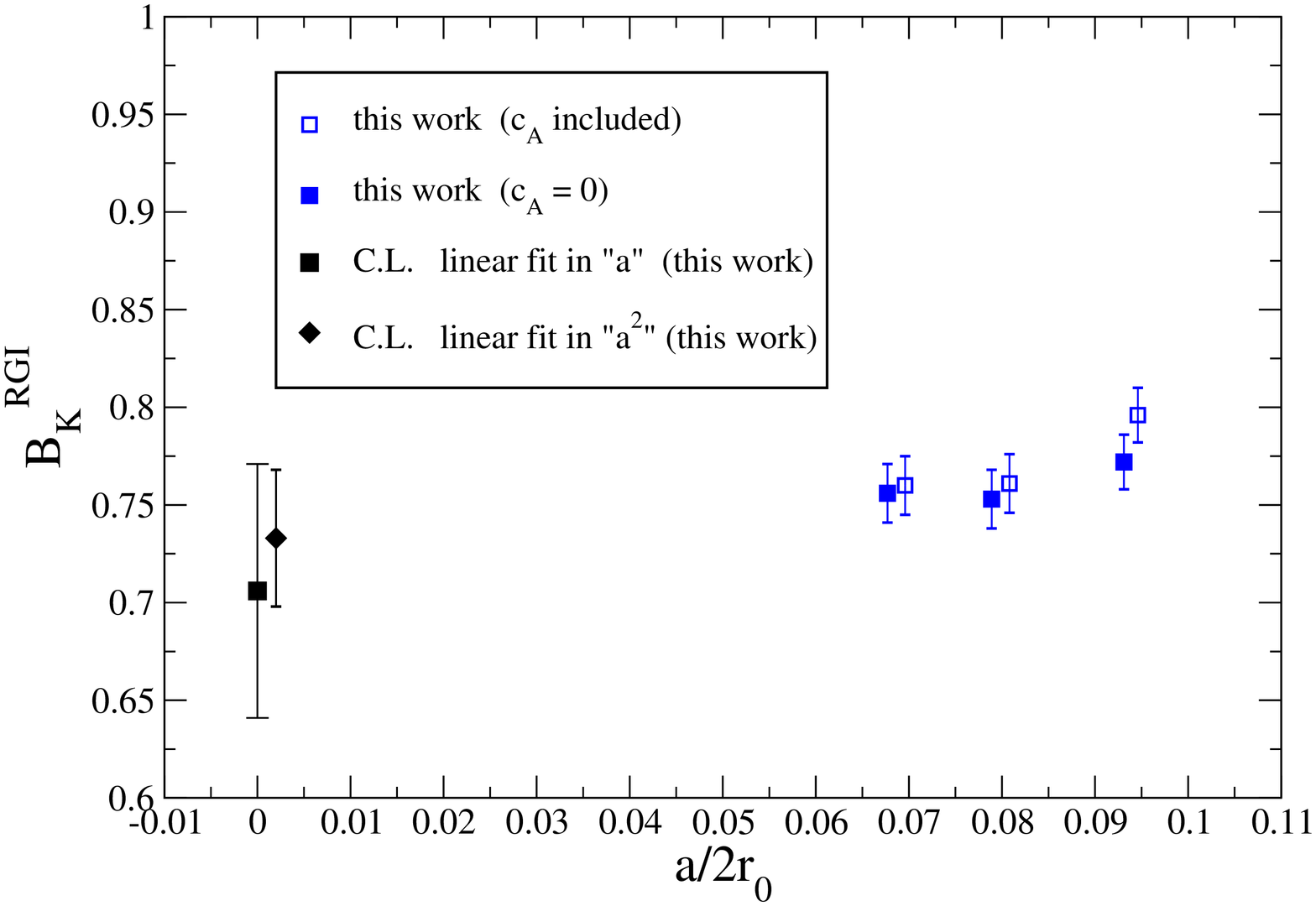}}
\caption[]{(a) Scaling behaviour of the $B_{\rm K}^{\rm RGI}$ results of 
refs.~\cite{Dimopoulos:2006dm,Dimopoulos:2007cn} for: 
(i) the $\pi/2$ case with a $c_A$ term (empty squares) and with $c_A = 0$ 
(filled squares); (ii) the $\pi/4$ case with a $c_A$ term (empty diamonds)
and with $c_A = 0$ (filled diamonds).
(b) Scaling behaviour of the $B_{\rm K}^{\rm RGI}$ results
from the present work with $c_A = 0$ (filled squares)
and with a $c_A$ term (empty squares).
Symbols have been shifted for clarity. See the text for
details on the continuum limit extrapolations (C.L.).}.
\label{BKscaling}
\end{center}
\end{figure}

In Table~\ref{tab:BK} we list our $B_{\rm K}^{\rm RGI}$ estimates. The scaling of these results is to be compared to the ones of refs.~\cite{Dimopoulos:2006dm,Dimopoulos:2007cn}. We remind the reader that the latter had been computed in two distinct frameworks: (i) the so-called $\pi/2$ case, consisting of a maximally twisted down quark and a non-twisted strange one and (ii)  the so-called $\pi/4$ case, in which down and strange flavours were in the same isospin doublet with a twist angle $\pi/4$. Both frameworks have quark flavours which are not maximally twisted and this implies that the results of refs.~\cite{Dimopoulos:2006dm,Dimopoulos:2007cn} are not improved.

In this respect, we wish to draw the reader's attention to a subtlety, related to the Symanzik-improvement term $a c_A \partial_0 P$ of the axial currents in the $B_{\rm K}$ denominator. In these earlier works cited above, the influence of this term in the scaling behaviour of $B_{\rm K}$ was carefully monitored. The result of this analysis is summarized in Fig.~\ref{BKscaling}(a). Note that $B_{\rm K}^{\rm RGI}$ scales much better when $c_A = 0$ (i.e. when the $B_{\rm K}$ denominator is not improved), than when it is tuned so as to eliminate $O(a)$ effects from the axial current. In ref.~\cite{Dimopoulos:2006dm} it was speculated that this could be due to some cancellation between the discretization effects of the $B_{\rm K}$ numerator and denominator, which is spoilt once the $c_A$ term is switched on. Whatever the reason, the data of Fig.~\ref{BKscaling}(a) clearly indicate that $O(a)$ effects influence significantly the scaling behaviour of $B_{\rm K}$. The continuum limit value shown in this figure is the final result quoted in ref.~\cite{Dimopoulos:2007cn}. It has been obtained by a combined fit, linear in $a/2r_0$, of the $\pi/2$ and $\pi/4$ data with a $c_A$ term in the denominator (which is the most conservative option).

In Fig.~\ref{BKscaling}(b) we show our results, plotted against the linear lattice spacing $a/2r_0$.
They are shown both for $c_A=0$ and for non-zero $c_A$, in order to
monitor the influence of the relevant counterterm.   
Note that due to automatic improvement, this is now an $O(a^2)$ effect. 
The compatibility between the two sets of results in Fig.~\ref{BKscaling}(b) indicates that, 
unlike the previous case, the $c_A$ term does not spoil the good scaling properties of  $B_{\rm K}$. 
Moreover, the almost flat scaling behaviour may suggest that the $O(a\Lambda_{\rm QCD})$ 
discretization effects of $Z_{VA+AV}$, as well as the $O(\mu^2 a^2)$ 
effects of the $K^0-\bar K^0$ matrix element, are small. 
In the light of this  we have performed two continuum limit extrapolations, one linear in
$a$ and another linear in $a^2$.
Both extrapolation results are shown in Fig.~\ref{BKscaling}(b)), for the $c_A=0$ data.

The linear fit of our $B_{\rm K}^{\rm RGI}$ data  with respect to $a$ gives the following continuum result:
$$B_{\rm K}^{\rm RGI} = 0.706(65) ~~~ (\chi^2/dof = 0.30) $$ 
while the linear fit with respect to $a^2$ gives:
$$~~B_{\rm K}^{\rm RGI} = 0.733(34) ~~~ (\chi^2/dof = 0.26)~. $$
Both results compare nicely with the published value: $B_{\rm K}^{\rm RGI} = 0.735(71)$ of ref.~\cite{Dimopoulos:2007cn}. Our error in the latter value is smaller because the extrapolation in $a^2$  is shorter.

\section{Conclusions}
\label{sec:conclusions}

In this work we have addressed a couple of issues related to the proposal of ref.~\cite{Frezzotti:2004wz} for the computation of $B_{\rm K}$ in a tmQCD setup, using a four-fermion operator made of two tm-type
quarks and two OS-type quarks, all maximally twisted. This ensures that the operator is multiplicatively renormalizable and automatically improved. The price to pay is that in this way the Kaon and anti-Kaon states of the weak matrix element $\langle \overline K^0 | O^{\Delta S = 2} | K^0 \rangle$ are composed of different lattice quark field combinations (tm and OS) and thus are non degenerate at finite lattice spacing.

We have performed a quenched study in order to address these issues. In particular, we find that the inclusion of the Clover term in the fermionic action significantly reduces the mass splitting between Kaon and anti-Kaon at finite lattice spacing. This result, already observed in an earlier work at a single lattice spacing, has been confirmed by us for several $\beta$-values. It should also be noted that in earlier studies, in which a Clover term was not included in the action,  this mass splitting, besides being much bigger than the one we measure, was found to be significantly affected by the choice of non-perturbative procedure for tuning the lattice theory to maximal twist. This does not appear to be the case once an $O(a)$ Symanzik-improved procedure, with a standard (untwisted) mass term, is used to tune the hopping parameter to its critical value. We also confirm that this splitting decreases as the continuum limit is approached.

The second objective of the present paper is the presentation of first results for $B_{\rm K}$, obtained in this tm-OS setup. We have compared our results to the earlier ones of refs.~\cite{Dimopoulos:2006dm,Dimopoulos:2007cn}, in which $B_{\rm K}$ had been computed in a tmQCD setup, without maximal twist on all flavours and thus without automatic $O(a)$-improvement. These results were rather sensitive to discretization errors, as revealed by the dependence of $B_{\rm K}$ on the Symanzik $O(a)$ term of the axial current at coarser lattice spacings.
In the present work, bare matrix elements are automatically
$O(a)$-improved, leaving us only with $O(a\Lambda_{\rm QCD})$ effects from
the renormalization constant of the four-fermion operator. Our results
display significantly better scaling, compatible with  an $O(a^2)$
behaviour.

\section*{Acknowledgments}
We gratefully acknowledge the participation of F.~Palombi at the early stages of this work and thank him for his help and suggestions. We also wish to thank R.~Frezzotti, G.~Herdoiza, C.~Pena, G.C.~Rossi, S.~Sint and R.~Sommer for discussions. We thank the theory groups at Milano-Bicocca and Tor Vergata for providing hospitality to members of our collaboration at various stages of this work. This work was supported in part by the EU Contract No. MRTN-CT-2006-035482, ``FLAVIAnet". All simulations have been performed on apeNEXT computers.

\bibliography{lattice}        

\end{document}